\documentclass{ws-ijgmmp}
\usepackage{amssymb}
\usepackage{amsmath}
\usepackage{amsfonts}
\usepackage{bm}
\def\be{\begin{align}}
\def\ee{\end{align}}
\newcommand\nn{\nonumber \\}
\newcommand\e{\mathrm{e}}

\begin{document}

\markboth{Nojiri, Odintsov, Faraoni}
{Alternative entropies and consistent black hole thermodynamics}

%%%%%%%%%%%%%%%%%%%%% Publisher's Area please ignore %%%%%%%%%%%%%%%
%
\catchline{}{}{}{}{}
%
%%%%%%%%%%%%%%%%%%%%%%%%%%%%%%%%%%%%%%%%%%%%%%%%%%%%%%%%%%%%%%%%%%%%

\title{ALTERNATIVE ENTROPIES AND CONSISTENT BLACK HOLE 
THERMODYNAMICS}

\author{SHIN'ICHI~NOJIRI}
\address{Department of Physics, Nagoya University\\ 
Nagoya 464-8602, Japan\\
Kobayashi-Maskawa Institute for 
the Origin of Particles and the Universe, Nagoya University\\
 Nagoya 464-8602, Japan\\
nojiri@gravity.phys.nagoya-u.ac.jp}

\author{SERGEI~D.~ODINTSOV}
\address{Instituci\'{o} Catalana de Recerca i Estudis 
Avan\c{c}ats (ICREA), Passeig Llu\'{i}s Companys 23\\
08010 Barcelona, Spain\\
Institute of Space Sciences (IEEC-CSIC), C. Can Magrans s/n\\ 
08093 Barcelona, Spain\\
odintsov@ieec.uab.es
}

\author{VALERIO FARAONI}
\address{Department of Physics \& Astronomy, Bishop's University, 2600 
College Street\\
Sherbrooke, Qu\'{e}bec, Canada J1M~1Z7\\
vfaraoni@ubishops.ca}

\maketitle

\begin{history}
\received{(Day Month Year)}
\revised{(Day Month Year)}
\end{history}

\begin{abstract}

While the Bekenstein-Hawking entropy is the unique notion of entropy that 
makes classical black hole thermodynamics consistent, alternative entropy 
notions (R{\'e}nyi, Tsallis, and generalized constructs) abound in the 
literature. We explore conditions under which they are part of a 
consistent horizon thermodynamics for certain classes of modified gravity 
black holes.  We provide examples in which black hole masses and 
temperatures going hand-in-hand with these alternative entropies coincide 
with their usual counterparts associated with the Bekenstein-Hawking 
entropy.

\end{abstract}


\section{Introduction}
\label{sec:1}
\setcounter{equation}{0}

Bekenstein's discovery \cite{Bekenstein:1973ur} of an entropy proportional 
to the area of a black hole event horizon was the first step  to black 
hole thermodynamics, no doubt a very insightful development in physics. An 
essential complement to Bekenstein's notion of entropy is Hawking's 
discovery \cite{Hawking:1975vcx} that black holes radiate scalar field 
quanta with a blackbody spectrum at a well-defined temperature, now called 
Hawking temperature. Black hole entropy and temperature were crucial 
ingredients in building a consistent  picture of 
black holes as thermal systems \cite{Bardeen:1973gs, Wald:1999vt, 
Carlip:2014pma}.

In the following we use the notation of Ref.~\cite{Wald}, in which the 
Lorentzian metric signature is ${-}{+}{+}{+}$, $\kappa^2\equiv 8\pi G$ 
where $G$ is Newton's constant, and units are adopted in which the speed 
of light $c$ and the reduced Planck constant $\hbar$ are unity.

For the prototypical Schwarzschild black hole with geometry 
\begin{equation}
\label{dS3BB}
ds^2= - \left( 1 - \frac{2GM}{r}\right) dt^2 + \frac{dr^2}{1-2GM/r} + r^2 
d\Omega^2_{(2)} 
\end{equation}
(where $M$ is the mass and $d\Omega_{(2)}^2 \equiv 
d\vartheta^2+ \sin^2 \vartheta \, d\varphi^2$ denotes the line 
element of the two-sphere of unit radius), the Bekenstein-Hawking entropy 
is proportional to the area $A=4\pi {r_\mathrm{h}}^2$  of the  
event horizon with radius  $r_\mathrm{h}=2GM$ \cite{Bekenstein:1973ur}.  
It is only when the internal energy $E$ and the temperature are identified 
with the black hole mass $M$ and the Hawking temperature 
\cite{Hawking:1975vcx} 
\begin{equation}
T_\mathrm{H}= \frac{1}{4\pi r_\mathrm{h}}= \frac{1}{8\pi 
GM} \,,
\end{equation}
respectively, that the area law of the Bekenstein-Hawking entropy 
\cite{Bekenstein:1973ur} is obtained. Then, the thermodynamical definition 
of entropy $dE= T d\mathcal{S} $, which here becomes a relation between 
$E,T$, and ${\mathcal S}$ mandatory for consistency, gives 
\begin{equation}
\label{S1}
d\mathcal{S}=\frac{dE}{T}=8\pi GM dM =d\left( 4 \pi G M^2 \right) 
\end{equation}
and 
\begin{equation}
\label{S2}
\mathcal{S}=4 \pi G M^2 + \mathcal{S}_0\,,
\end{equation}
with $\mathcal{S}_0$  an integration constant. When there is no black hole 
$M$ and $\mathcal{S}$ 
vanish simultaneously, which fixes the integration constant 
$\mathcal{S}_0$ 
to zero, hence the resulting area law 
\begin{equation}
\label{S3}
\mathcal{S}= \frac{A}{4G} = \frac{\pi {r_\mathrm{h}}^2}{G} 
\end{equation} 
is a consequence of assuming $E=M$ and $T=T_\mathrm{H}$.

As done in our previous work~\cite{Nojiri:2021czz}, it is legitimate to 
ask whether it is necessary to assume $E=M$ and $T=T_\mathrm{H}$. The 
first assumption is backed  by a {\em gedankenexperiment} 
introduced in~\cite{Nojiri:2021czz} and based on an 
infalling spherical dust shell of mass $M$ and initially large  
radius. The Birkhoff theorem \cite{Wald} guarantees that the spacetime 
exterior to the shell is Schwarzschild, where the constant $M$  in the 
line element~(\ref{dS3BB}) 
is now the shell mass. By the same theorem, the interior metric is 
the Minkowski one. A black hole forms when the shrinking shell crosses 
its Schwarzschild radius $r_\mathrm{h}$. As the spacetime  
geometry remains asymptotically flat during collapse, the mass $M$ in the 
line element~(\ref{dS3BB}) remains the shell mass  and coincides with the 
system's energy $E=M$. The latter is conserved during  collapse 
because the exterior static geometry remains Schwarschild with 
Schwarzschild mass $M$ (the initial mass of the shell). The  
spherical symmetry prevents the emission of gravitational waves 
(quadrupole waves in general relativity) carrying away energy.

Next, should one necessarily identify the temperature with the Hawking 
temperature? The calculation of the Hawking temperature, which neglects 
backreaction on the black hole, can only  proceed 
after the horizon geometry is fixed, which allowed Hawking to 
compute $T_\mathrm{H}$  as the quantity appearing in the thermal 
emission of  (scalar) radiation. Suppose that  
the black hole is located in a thermal bath at 
temperature $T$: then, after  a transient, thermal equilibrium is 
established  between Hawking radiation 
and heat bath at a final common  temperature  $T=T_\mathrm{H}$. The 
heat bath is the analogue of a thermometer measuring the 
black hole (Hawking) temperature. 

Let us come now to possible variations. Different notions of entropy have 
appeared in recent literature, beginning with the Tsallis entropy 
\cite{Tsallis:1987eu,Ren:2020djc,Nojiri:2019skr} and the R{\'e}nyi entropy 
\cite{renyi, Czinner:2015eyk,Tannukij:2020njz, 
Promsiri:2020jga,Samart:2020klx}), and continuing with the Sharma-Mittal 
\cite{SayahianJahromi:2018irq}, Barrow \cite{Barrow:2020tzx}, and 
Kaniadakis \cite{Kaniadakis:2005zk,Drepanou:2021jiv} entropies. Another 
entropy definition arises in the context of non-extensive statistical 
mechanics applied to Loop 
Quantum Gravity \cite{Majhi:2017zao,Czinner:2015eyk,Mejrhit:2019oyi, 
Liu:2021dvj}. In spite of the differences, one recognizes four  
properties common to all these entropies: 
positivity, monotonicity, Bekenstein-Hawking limit, and  generalized third 
law.

To begin with, these entropies are always positive, as is the 
Bekenstein-Hawking entropy~(\ref{S2}). In fact, $\e^\mathcal{S}$ is the 
number of states, or their volume, and $\e^\mathcal{S}>1$ (indeed, 
$\e^\mathcal{S} \gg 1$). 
Second, all these entropies strictly increase with   
the Bekenstein-Hawking entropy $\mathcal{S}$. Third, these alternative 
entropies admit a limit in which they become the Bekenstein-Hawking 
entropy~(\ref{S2}). Finally, these entropies go to zero when the 
Bekenstein-Hawking entropy ${\mathcal S}$ is zero. The 
standard statistical mechanics of closed systems in thermal equilibrium 
interprets  $\e^\mathcal{S}$ as the number of states and, accordingly, 
if the ground, or vacuum, state is unique $\mathcal{S}$ vanishes 
when the temperature hits zero. The Bekenstein-Hawking 
entropy $\mathcal{S}$ is very different in this sense because it diverges 
when $T \to 0$ and vanishes asymptotically at infinite temperatures. 
It seems natural that all generalized  entropies should go to zero as 
$\mathcal{S} \to 0$. 

Adopting these four properties, in our previous work 
\cite{Nojiri:2022aof} we proposed two new definitions of 
generalized entropy containing, respectively, six or 
three parameters and reproducing, in appropriate limits, the generalized 
entropy notions of the literature mentioned above.  Could these  
entropies replace the Bekenstein-Hawking entropy for non-Schwarzschild  
black hole solutions of  theories of gravity alternative to general 
relativity? There is plenty of such alternative gravity theories, 
and of their solutions (recently reviewed in \cite{Faraoni:2021nhi}), in the 
literature. 
This question is 
answered affirmatively in the following, where we restrict to 
static and spherical black holes. 

The thermodynamical energy of such black holes in modified gravity is 
discussed in the next section, while their temperature going hand-in-hand 
with an entropy alternative to the Bekenstein-Hawking one is the subject 
of Sec.~\ref{sec:3}. Section~\ref{sec:4} proposes specific models in 
featuring the Tsallis, R{\'e}nyi, or other generalized entropy, while 
conclusions are drawn in  Sec.~\ref{sec:5}.

\section{Thermodynamic energy associated with alternative entropy}
\label{sec:2}

Consider static, spherical, and asymptotically flat spacetimes in 
alternative  
theories of gravity, described by the line element
\begin{align}
\label{metric}
ds^2 = g_{\mu\nu} dx^{\mu} dx^{\nu} 
= -\e^{2\nu(r)} dt^2 + \e^{2\lambda(r)} dr^2
+ r^2 d\Omega_{(2)}^2 \, .
\end{align}
Asymptotic flatness corresponds to $\lim_{r\to +\infty} 
\lambda(r )=0$ and,  normalizing the time 
coordinate $t$, to $\lim_{r\to+\infty} \nu( r )=0$.

Let us begin with Einstein gravity and interior solutions inside matter, 
using the Tolman-Oppenheimer-Volkov (TOV) equation to discuss the mass. 
The time-time component of the Einstein equations reads
\begin{align}
\label{EinTOV0}
 - \kappa^2 \rho = \frac{1}{r^2} \left( r \e^{-2\lambda} - r \right)' \,,
\end{align}
where $\rho$ is the energy density and a prime denotes 
differentiation with respect to $r$. The mass is identified by writing  
\begin{equation}
\e^{-2\lambda} \equiv 1 - \frac{\kappa^2 m(r)}{4\pi r} \,,
\end{equation}
from which it follows that $4\pi r^2 \rho = m'(r) $ and, integrating,  
\begin{align}
\label{TOV1}
m(r) =4 \pi \int^r_{0} r^{\prime 2} \rho (r') dr'  + m_0 \,,
\end{align}
where $m_0$ is an integration constant. In a (compact)  
star, the solution must be regular at the centre. Moreover,  we impose 
\begin{align}
\label{consin}
\lambda\to 0 \, , \quad\quad 
\lambda'(r) =\frac{m-rm'}{r(r-2m)} \xrightarrow{} 0 
\end{align} 
as $r\to0$ to avoid a 
conical singularity, fixing $m_0 =0$ and 
\begin{align}
m(r)=4 \pi \int^ r_{0} r^{\prime 2} \rho (r') dr'  \,.
\label{eq:massfunc}
\end{align}
If the geometry is asymptotically Schwarzschild, the mass is
\begin{align}
\label{mass}
M=m(r\to\infty)=4 \pi \int_0^\infty d r \, r^2 \rho(r) \,.
\end{align}
A different situation occurs if there is a 
central singularity, as in  black holes: in this case  the integration 
constant $m_0$ is chosen so that
\begin{align}
\label{massB}
M=m(r=\infty)=4 \pi \int_0^\infty d r\, r^2 \rho(r) + m_0 
\end{align}
and now  $m(r=\infty)$ is not the total mass, which is instead defined by 
\begin{eqnarray}
\bar{M} &=& \int d^3 x \, \sqrt{\gamma} \, \rho(r) 
=4 \pi \int_0^\infty  \rho(r) r^2 \e^{\lambda(r)} d r \nonumber\\
&=& 4 \pi \int_0^\infty \rho(r) r^2 \left[1-\frac{2 G m(r)} 
r\right]^{-1/2}  d r \nonumber\\
& = & 4 \pi \int_0^{\infty} dr \, \rho(r) r^2 \left[1 + \frac{G m(r)} r - 
\frac{3G^2 m^2(r)}{r^2} + \mathcal{O} \left( G^3 \right) \right] 
\,, \label{barM}
\end{eqnarray} 
where $\gamma$ denotes the determinant of the three-dimensional 
Riemannian metric 
\begin{align}
\label{induced3metric}
\gamma_{\ell m} \, d x^{\ell} d x^m =\e^{2 \lambda} d r^{2}+r^{2} 
d\Omega_{(2)}^2 \, .
\end{align}
Of course, we wish to compare this mass with the Newtonian mass. Let us 
use  $G$ instead of $\kappa^2 \equiv 8\pi G$. The difference between the 
Schwarzschild mass $M$ in~(\ref{mass}) and the 
total mass~$\bar{M}$ is the gravitational binding energy of the 
spherical object $ E_{\mathrm{B}}=M-\bar{M} $.  We interpret the second 
term in the last line of Eq.~(\ref{barM}) as the  
Newtonian gravitational potential energy 
\begin{align}
\label{NP}
 - 4 \pi G \int_0^\infty dr \, \rho(r) \, r^2 \frac{m(r)} r = - 
\frac{G}{2} \int 
dV \int dV' \, \frac{\rho\left(\bm r\right) \rho\left(\bm 
r'\right)}{\left| 
\bm r - \bm r' \right|} \, ,
\end{align}
where $dV$ and $dV'$ are three-dimensional volume elements and the 
general-relativistic nonlinear corrections are identified by 
$G^2$ and higher powers of $G$.

For a singular black hole, if the  geometry is asymptotically 
Schwarzschild the integration constant $m_0$ in~(\ref{TOV1}) can be 
fixed by imposing that $m(r\to \infty)=M$. For  the  
vacuum Schwarzschild black hole, the energy density $\rho$ is 
identically zero and $m_0=M$, which can be regarded as 
the contribution from a Dirac delta centered at $r=0$ 
\cite{Heinzle:2001bk,Steinbauer:2006qi}. As a result,  the mass obtained 
here coincides with the usual ADM mass.

Moving from general relativity to modified gravity, one can still write 
the time-time component of the field equations as 
\begin{align}
\label{MGTOV}
 - \kappa^2 \rho_\mathrm{eff} = \frac{1}{r^2} \left( r \e^{-2\lambda} - r 
\right)' \,,
\end{align}
but now $ \rho_\mathrm{eff}$ is an effective energy density obtained by 
writing the field equations as effective Einstein equations containing 
an effective stress-energy tensor (made of the non-Einsteinian 
gravitational terms) in their right-hand sides. This procedure yields the 
effective mass
\begin{align}
\label{MDmassfunc}
m_\mathrm{eff}(r)=4 \pi \int^r_0dr' r^{\prime 2} \rho_\mathrm{eff} 
(r')\,,
\end{align}
that we interpret as the mass acted upon by the attractive force at 
radius $r$. For example, consider $F(R)$ gravity with action 
\begin{align}
\label{FR}
S_{F(R)} =\frac{1}{2\kappa^2} \int d^4 x \sqrt{-g} \, F(R) + 
S^\mathrm{(matter)}\,,
\end{align}
where  $F(R)$ depends non-linearly on the Ricci scalar $R$ and 
 $g$ is the determinant of the spacetime metric $g_{\mu\nu}$, while 
$ S^\mathrm{(matter)}$ denotes the matter part of the action. Using the 
notation 
$F(R)=R+f(R)$ and $f_R(R) \equiv df(R)/dR$,  the time-time  field equation 
defines the total  ({\em i.e.}, matter plus effective) energy density $ 
\rho_\mathrm{eff} = \rho + \rho_{F(R)} $, where 
\begin{align}
\rho_{F(R)} \equiv &\, \frac{1}{\kappa^2} \left\{ 
 - \frac{f}{2} - \e^{- 2 \lambda} \left[
\nu'' + \left(\nu' - \lambda'\right)\nu' + \frac{2\nu'} r\right] f_R 
\right.
\nonumber\\
&\left. 
+ \e^{ -2\lambda} \left[ f_R'' + \left( - \lambda' + \frac{2} r \right) 
f_R' \right] \right\} \,,
\end{align}
which produces the effective total mass  
$\bar{M}_\mathrm{eff}$ of Eq.~(\ref{barM}). The latter contains  
contributions from both matter and gravity since 
\begin{align}
\label{barMeff}
\bar{M}_\mathrm{eff} = \int d^3 x \, \sqrt{\gamma}\, \rho_\mathrm{eff}(r) 
= \int d^3 x \, \sqrt{\gamma} \left( \rho + \rho_{F(R)} \right) \, . 
\end{align}
The leading correction in the binding energy of Eq.~(\ref{NP}) 
is 
\begin{align}
\label{bindingeff}
E_{\mathrm{B,eff}} = - G \int dV \int dV' \, \frac{\left[ \rho\left(\bm 
r\right) + \rho_{F(R)}\left(\bm r\right) \right]
\left[ \rho\left(\bm r'\right) + \rho_{F(R)}\left(\bm r'\right) 
\right]}{ \left| \bm r - \bm r' \right|} + \cdots 
\end{align}
and comprises contributions from the interactions between matter and 
gravitational 
energy in $F(R)$ gravity,  between the gravitational 
energy densities of $F(R)$ gravity, and from the 
self-interaction of matter via the gravitational force. 
Accordingly, we interpret $M_\mathrm{eff}\equiv m_\mathrm{eff}\left( 
r\to\infty \right)$ as the total mass-energy of the system while 
$m_\mathrm{eff}(r)$ is the mass-energy contained in a 2-sphere of radius 
$r$.

A black hole in modified gravity can have horizon 
radius $r_\mathrm{h}$ different from $2G M_\mathrm{eff}  \equiv 
2 m_\mathrm{eff} \left(r\to 
\infty\right) $. In general, if one decides to use $M_\mathrm{eff}$ as the 
internal energy and  $\mathcal{S}=4\pi {r_\mathrm{h}}^2 /4$ as the black 
hole entropy, then the temperature 
\begin{align}
\label{temperature}
\frac{1}{T}=\frac{d\mathcal{S}}{dM_\mathrm{eff}}
\end{align}
does not coincide with the Hawking temperature $T_\mathrm{H}$. 
Alternatively, it is possible to do thermodynamics using the Hawking 
temperature in conjunction with  the entropy 
\begin{align}
\label{entropy}
d\mathcal{S} = \frac{dM_\mathrm{eff}}{T_\mathrm{H}}
\end{align}
instead of the Bekenstein-Hawking entropy. However, in general,  
$M_\mathrm{eff} \neq m_\mathrm{eff}\left( 
r_\mathrm{h} \right)$ and the difference $M_\mathrm{eff} - 
m_\mathrm{eff}\left( r_\mathrm{h} \right)$ could correspond to the energy 
outside of the horizon. Restricting to the thermodynamics of this black 
hole, one would  
identify  $m_\mathrm{eff}\left( r_\mathrm{h} 
\right)$ with its internal energy and modify  
Eq.~(\ref{entropy}) as  
\begin{align}
\label{entropy2}
d\mathcal{S}_\mathrm{bh} = \frac{d m_\mathrm{eff}\left( r_\mathrm{h} 
\right)}{T_\mathrm{H}}\,.
\end{align}

\section{Temperature associated with alternative entropy}
\label{sec:3}

Let us adopt the notation 
\begin{align}
\label{ss1}
h(r) \equiv \e^{2\nu(r)} \, , \quad \quad h_1(r) \equiv \e^{-2\lambda(r)} 
\,,
\end{align}
where the vanishing of $h(r)$ locates the black hole event horizon. We are 
going to show that, in general, if $h_1(r)$ does not vanish simultaneously 
with $h(r)$, the spacetime curvature diverges on the surface $h(r)=0$. If, 
instead, $h_1(r)$ vanishes simultaneously with $h(r)$, the curvature 
remains finite and the surface $ h_1(r)=h(r) =0$ is an event horizon.

The proof of these statements hinges on the Kretschmann scalar, the 
square of the Ricci 
tensor, and the Ricci scalar
\begin{align}
\label{inv33}
R_{\mu \nu \rho \sigma} R^{\mu \nu \rho \sigma} =&\, \frac {1}{4h^4 
r^4}\left[ 4 r^4 h''^2 h^2 {h_1}^2
+4 r^4hh_1 h' h'' \left( h'_1 h - h' h_1 \right)  +  \left( h'^2 {h_1} 
r^2\right)^2  \right. \nonumber\\
& \left. -2 r^4 h'^3h_1h'_1h  + \left(r h {h'}\right)^2 \left( {h_1'}^2 
r^2+8  {h_1} \right)^2   \right.\nonumber\\
& \left. +8 \, h^4 
\left(  r^2 {h'_1}^2 + 2\, \left( 1- h_1 \right)^2 \right) \right] \,, \\
&\nonumber\\
R_{\mu \nu } R^{\mu \nu }=&\, \frac {1}{8h^4 r^4}\left[ 4 r^4 h''^2h^2 
{h_1}^2+4 h 
\left[h \left( rh'_1 +2h_1 \right)h'- r h'^2h_1 +2\ h^2h'_1 \right] 
r^3h_1 h'' \right. \nonumber\\
& + r^4h'^4{h_1}^2  + r^2h^2 \left( 12 {h_1}^2+{h'_1}^2 r^2 \right)h'^2 -2 
r^3h h_1 \left( 
rh'_1 +2 h_1 \right) h'^3 \nonumber\\
& \left. +4r h^3 \left( 2h'_1rh_1 -4h_1 +4 {h_1}^2+ {h'_1}^2 r^2 \right) 
h' \right. \nonumber\\
& \left. 
+4h^4 \left(3 {h'_1}^2 r^2 +4r \left( h_1-1 \right)h'_1 +4\left(h_1 
-1 \right)^2 \right) \right] \,,\\
&\nonumber\\
R=&\, \frac{2 h'' h_1 h r^2- r^2h_1  h'^2 +r h' h \left( rh'_1+ 4h_1   
\right) +4h^2 \left( h_1 + rh'_1  -1 \right) }{2 h^2 r^2 }\,.
\end{align}
The denominators of these algebraic curvature invariants contain positive 
powers of $h(r)$ that make these invariants diverge in the limit $h \to 
0$. If $h_1(r)$ vanishes simultaneously with $h(r)$, 
the  invariants~(\ref{inv33}) remain finite at the 
roots of $ h_1(r) = h(r) =0$. In fact, if  
$h_1(r) = h_2(r) h(r) $ and   $h_2 \neq 0 $ and is regular 
at  the roots of $h(r)=0$, substituting  $h_1 = h_2 h$ in 
Eqs.~(\ref{inv33}) gives  
\begin{align}
\label{invariants2}
R_{\mu \nu \rho \sigma} R^{\mu \nu \rho \sigma}
=&\, {h''}^2 {h_2}^2 + h' h_2 h_2'  + \left(  \frac{ {h'} {h_2'} }{2} 
\right)^2  + 
\left( \frac{2 {h_2} {h'}}{r} \right)^2 \nonumber\\
& + \frac{2\left[ r^2 \left( h h_2'  + h' h_2  \right)^2  
+ 2 \left( {h_2}^2 h^2 - 1 \right)^2 \right] }{r^4} \, , \\
&\nonumber\\
\label{invariants3}
R_{\mu \nu } R^{\mu \nu } =&\, \frac{{h''}^2 {h_2}^2}{2} + \frac{h'' h' 
h_2 h_2'}{2} + \frac{{h_2}^2 h' h''  } r + \frac{ h_2 h^2  \left( h h_2'  
+  h_2 h'\right) }{r} 
+ \frac{3 {h_2}^2 {h'}^2 }{2r^2} \nonumber\\
& + \frac{{h'}^2 {h_2}^2 }{8} + \frac{ h_2 h'  
\left( h h_2' + h_2 h' \right)}{r^2} \nn
& - \frac{2 {h_2}^2 h' }{r^3} + \frac{2h {h_2}^2 h' }{r^3} + \frac{h  
{h_2}^2 h' }{2r}  + \frac{ h_2  {h'}^2 h_2'} r \,\nonumber\\
& \times \, 
 \frac{3 \left( h h_2' + h_2 h' \right) r^2
+4r \left( h h_2 -1 \right) \left( h h_2' + h_2 h' \right) +4\left(1-h h_2 
 \right)^2}{2r^4} \, , 
\end{align}
and
\begin{align}
\label{invariants1}
R=&2\, h_2  h'' + \frac{2 \, r h_2 h'}{r} + \frac{ h' h_2' }{2} + 
\frac{2\left[ h_1+   r \left( h h_2'  +  h_2 h'\right)  - 1\right]}{r^2} 
\,;
\end{align}
an inspection shows that these invariants are finite when $h(r)=0$.

Since both $h_1(r)$ and $h(r)$ vanish at the horizon, one can write 
$h_1(r) =  \e^{-2\lambda(r)}$ using $m(r)$ and the horizon radius takes 
the form  
\begin{align}
\label{horizon}
r_\mathrm{h}= \frac{\kappa^2 m(r_\mathrm{h})}{4\pi}= 2G m(r_\mathrm{h}) 
\, .
\end{align}
The corresponding Hawking temperature is obtained geometrically. Near the 
horizon the radial coordinate is $r \equiv r_\mathrm{h} + \delta r$ and 
\begin{eqnarray}
\e^{-2\lambda} &=& h_1 = h h_2  = \frac{C\left( r_\mathrm{h} 
\right) \left( r - r_\mathrm{h} \right)   }{r_\mathrm{h}} \,,\\ 
\e^{2\nu} &=& h = \frac{h_1}{h_2} = \frac{C\left( r_\mathrm{h} 
\right) \left( r - r_\mathrm{h} \right)  }{h_2 \left( r_\mathrm{h} \right) 
r_\mathrm{h}}  \,,
\end{eqnarray} 
where $C\left( r_\mathrm{h} \right)\equiv 1 - m'\left( r_\mathrm{h} 
\right)$. After a  Wick rotation of the time coordinate $t\to i\tau$, the 
line element~(\ref{metric}) in the vicinity of the horizon  becomes 
\begin{align}
\label{TH1}
ds^2 \simeq \frac{C\left( r_\mathrm{h} \right) \delta r}{h_2 \left( 
r_\mathrm{h} \right) r_\mathrm{h}} \, d\tau^2 
+ \frac{r_\mathrm{h}}{C\left( r_\mathrm{h} \right) \delta r} \, d(\delta 
r)^2 
+ r_\mathrm{h}^2 \, d\Omega_{(2)}^2 \,.
\end{align}
This form can be simplified by passing to a new radius  $ \rho$ 
defined in differential form by $ d\rho = d 
\left( \delta r\right) \sqrt{ \frac{r_\mathrm{h}}{C\left( r_\mathrm{h} 
\right) \delta r}} $
or in finite form by  
\begin{align}
\label{TH2}
\rho = 2 \, \sqrt{ \frac{r_\mathrm{h} \, \delta r}{C\left( r_\mathrm{h} 
\right)}} \quad\quad
\mbox{and} \quad\quad \delta r= \frac{C\left( r_\mathrm{h} \right) 
\rho^2}{4 r_\mathrm{h}} \,.
\end{align}
Near the horizon, the line element~(\ref{TH1}) assumes the form 
\begin{align}
\label{TH3}
ds^2 \simeq \frac{C\left( r_\mathrm{h} \right)^2}{ 4 h_2 \left( 
r_\mathrm{h} \right) r_\mathrm{h}^2} \, \rho^2d\tau^2
+ d\rho^2 + r_\mathrm{h}^2 \, d\Omega_{(2)}^2 \,.
\end{align}
It is necessary to impose that the Euclidean time coordinate $\tau$ is 
periodic to prevent conical singularities near $\rho = 0$ in the 
 Euclidean space    generated by the Wick rotation, 
\begin{align}
\label{TH4}
\frac{C\left( r_\mathrm{h} \right) \tau}{ 2 r_\mathrm{h} \sqrt{h_2 \left( 
r_\mathrm{h} \right)}} 
\simeq \frac{C\left( r_\mathrm{h} \right) \tau}{ 2 \, r_\mathrm{h} \, 
\sqrt{h_2 \left( r_\mathrm{h} \right)}} + 2 \, \pi 
\end{align}
and then the temperature corresponds to inverse of the period $t_*$ of the 
Euclidean time. In finite temperature field theory, 
the Euclidean path integral 
formulation gives  
\begin{align}
\label{TH5}
\int \left[ D\phi \right] \e^{ \int_0^{t_*}  L(\phi)} dt 
= \, \mathrm{Tr}\left( \e^{-t_* H} \right) = \, \mathrm{Tr} \left( \e^{ - 
\frac{H}{T} } \right) 
\end{align}
and Schwarzschild black hole is endowed with the corresponding temperature 
\begin{align}
\label{TH6}
T = \frac{C\left( r_\mathrm{h} \right)}{4\pi r_\mathrm{h}\sqrt{h_2 \left( 
r_\mathrm{h} \right)}} 
= \frac{C\left( r_\mathrm{h} \right)}{8\pi G 
m_\mathrm{eff} \left( r_\mathrm{h} \right) \sqrt{h_2 \left( r_\mathrm{h} 
\right)} } 
= \frac{C\left( r_\mathrm{h} \right) T_\mathrm{H}}{\sqrt{h_2 \left( 
r_\mathrm{h} \right)}} \,,
\end{align}
where the Hawking temperature is now 
\begin{equation}
T_\mathrm{H}\equiv \frac{1}{8\pi G m_\mathrm{eff} \left( r_\mathrm{h} 
\right)} \,.
\end{equation}
In general,  $T$ deviates from the Hawking 
temperature by the factor\footnote{Here we call Hawking temperature 
the quantity $T_\mathrm{H} \equiv \frac{1}{4\pi r_\mathrm{h}}$ without the 
factor $\frac{C \left( r_\mathrm{h} \right)}{\sqrt{h_2 \left( r_\mathrm{h} 
\right)}}$. 
The temperature~(\ref{TH6}) is given by the surface gravity 
$\kappa$, $T=\frac{\kappa}{2\pi}$, as in the standard formulation.}  
$\frac{C \left( r_\mathrm{h} \right)}{\sqrt{h_2 
\left( r_\mathrm{h} \right)}}$, which cannot be absorbed into a  
rescaling of time because (as mentioned after 
Eq.~(\ref{metric})) we have fixed the scale so that 
\begin{equation}
h\left( r \to \infty 
\right) = h_2\left( r \to \infty \right) h_1\left( r \to \infty \right) 
=\e^{2\nu\left( r \to \infty \right) } =1 \,.
\end{equation}
Hawking radiation is obtained from the near-horizon geometry and the 
thermal distribution of the emitted radiation can correspond to the 
new temperature~(\ref{TH6}). Placing this black hole in a thermal bath in 
equilibrium at the temperature $T$,  the latter  equals  
the temperature~(\ref{TH6}), which is then identified with the black hole 
temperature.

As seen in the previous section, identifying $m_\mathrm{eff} 
\left( r_\mathrm{h} \right)$ with the black hole internal energy, 
Eq.~(\ref{entropy2}) implies that 
\begin{align}
\label{entropy3B}
\mathcal{S}_\mathrm{bh} = \int \frac{d m_\mathrm{eff} \left( r_\mathrm{h} 
\right)}{T}\, .
\end{align}
Integrating the  field equations of a certain gravitational theory, 
multiple 
integration constants $c_i$ $\left(i=1, \, \cdots \, , N  \right)$ appear.  
$N$ is larger in theories with higher order field equations, corresponding 
to more degrees of freedom.  For example, in general relativity the mass 
$M$ of the 
Schwarzschild black hole appears in the metric coefficients  
$ \e^{2\nu}=\e^{-2\lambda}= 1 - 2M/r $ as 
an integration constant. The number $N$ of integration constants depends 
on the theory and  $\lambda(r)$, $\nu(r)$ 
(and, therefore,  $m(r)$, $h(r)$, and $h_{1,2}(r)$) depend on the 
$c_i$'s.  Equation~(\ref{horizon}) is 
solved for $ r_\mathrm{h} \left(c_i\right)$ as a function of 
these integration constants. For the usual Schwarzschild black hole 
of general relativity,  this relation gives the familiar Schwarzschild 
radius 
$r_\mathrm{h}=2M$. Other quantities are obtained as functions 
of $c_i$, such as $ h_2\left(r=r_\mathrm{h} \left(c_i\right); 
c_i\right)$, {\it etc}. Equation~(\ref{horizon}) 
yields $m\left(r_\mathrm{h} 
\right)=m\left( r=r_\mathrm{h} \left(c_i\right); 
c_i\right)=\frac{r_\mathrm{h} \left(c_i\right)}{2G}$, which implies that  
the  
$c_i$'s can be parametrized using a single parameter $\xi$,  
$c_i=c_i(\xi)$. For example, for 
the Reissner-Nordstr\"{o}m black hole one can fix the electric charge and 
choose the mass to be this single parameter $\xi$ (which is equivalent to  
using the charge-to-mass ratio as a parameter). 
Proceeding in this way,~(\ref{TH6}) is used to turn  Eq.~(\ref{entropy3B}) 
in the form 
\begin{align}
\label{entropy3c}
\mathcal{S}_\mathrm{bh} = \frac{1}{2G} \int d\xi\, \frac{\left[ 4\pi 
r_\mathrm{h} \left( c_i \left(\xi\right) \right) 
\sqrt{ h_2\left(r=r_\mathrm{h} \left(c_i\left(\xi\right) \right); 
c_i\left(\xi\right)\right) } \right]}
{1 - \left. \frac{\partial m\left(r; c_i\left(\xi\right) \right)}{\partial 
r}\right|_{r= r_\mathrm{h} \left( c_i \left(\xi\right) \right)}}
\sum_{i=1}^N \frac{\partial r_\mathrm{h} \left(c_i\right)}{\partial c_i} 
\, \frac{\partial c_i}{\partial \xi}\, .
\end{align}
With the choice $\xi=r_\mathrm{h}$, Eq.~(\ref{entropy3c}) reduces to 
\begin{align}
\label{entropy3d}
\mathcal{S}_\mathrm{bh} = \frac{1}{2G} \int_0^{r_\mathrm{h}} d\xi \, 
\frac{\left(4\pi \xi \sqrt{ h_2\left(r=\xi; c_i\left(\xi\right)\right) } 
\right) }{1 - \left. \frac{\partial m\left(r; c_i\left(\xi\right) 
\right)}{\partial r}\right|_{r= \xi}} \,,
\end{align}
where the integration constant is determined by the condition 
$\mathcal{S}_\mathrm{bh}=0$ at $r_\mathrm{h} =0$. For the Schwarzschild 
black hole with $h_2(x)=1$, $m=M=$~const., one re-obtains the 
Bekenstein-Hawking entropy~(\ref{S3}). If, instead, $h_2\left( r \to 
r_\mathrm{h} \right)$ gives a 
non-trivial contribution, the entropy $\mathcal{S}_\mathrm{bh}$ can 
differ from the Bekenstein-Hawking entropy $\mathcal{S}_\mathrm{BH}$.

Equation~(\ref{entropy3d}) tells us that 
\begin{align}
\label{entropy5}
\frac{h_2\left(r=r_\mathrm{h}; c_i\left(r_\mathrm{h}\right)\right) } 
{\left(1 - \left. \frac{\partial m\left(r; c_i\left(r_\mathrm{h}\right) 
\right)}{\partial r}\right|_{r= r_\mathrm{h}}\right)^2}
= 16G^2 \Big[ \mathcal{S}_\mathrm{bh}' \left(A\right) \Big]^2 \,;
\end{align}
then, for certain expressions of the general entropies, we can find the 
corresponding form of 
\begin{equation}
\frac{h_2\left(r=r_\mathrm{h}; 
c_i\left(r_\mathrm{h}\right)\right) } 
{\left(1 - \left. \frac{\partial m\left(r; c_i\left(r_\mathrm{h}\right) 
\right)}{\partial r}\right|_{r= r_\mathrm{h}}\right)^2} \,.
\end{equation}

By now, several alternative notions of entropy have been introduced, with 
various motivations. The first was the R{\'e}nyi entropy unrelated to 
statistics introduced in 1960 \cite{renyi} to  
quantify the amount of  information. It was used in many works, 
{\em e.g.}, \cite{Czinner:2015eyk,Tannukij:2020njz, Promsiri:2020jga, 
Samart:2020klx}, it  contains a single parameter $\alpha$, and is simply 
\begin{align}
\label{RS1}
\mathcal{S}_\mathrm{R}=\frac{1}{\alpha} \ln \left( 1 + \alpha 
\mathcal{S} \right) 
\end{align}
where $\mathcal{S}$ is the Bekenstein-Hawking 
entropy.

Another widely studied possibility is the Tsallis entropy 
\cite{Tsallis:1987eu,Ren:2020djc,Nojiri:2019skr} 
\begin{align}
\label{TS1}
\mathcal{S}_\mathrm{T} = \frac{A_0}{4G} \left( \frac{A}{A_0} 
\right)^\delta 
\end{align}
originating in the 
non-extensive  statistics of physical systems with long range 
interactions, where the Boltzmann-Gibbs entropy becomes  inadequate
because the partition function diverges. 
Here the constant $A_0$ has the dimensions of a length squared and 
the 
dimensionless parameter $\delta$ measures the non-extensivity. Clearly, 
as $\delta \to 1$, $\mathcal{S}_\mathrm{T}$ reduces to  the 
Bekenstein-Hawking entropy.

The Tsallis entropy is used to define the more complicated  Sharma-Mittal 
entropy 
\cite{SayahianJahromi:2018irq}
\begin{align}
\label{SM}
\mathcal{S}_\mathrm{SM} = \frac{1}{R}\left[ \left(1 + \delta 
\, \mathcal{S}_\mathrm{T} \right)^{R/\delta} - 1 \right] \,,
\end{align}
with two 
phenomenological parameters $R$ and $\delta$ to be determined by  
experiment. This entropy construct interpolates between  the R{\'e}nyi 
and the Tsallis entropies.

Another construct, the Kaniadakis entropy 
\cite{Kaniadakis:2005zk,Drepanou:2021jiv} extends the familiar   
Boltzmann-Gibbs entropy to relativistic  systems 
\cite{Kaniadakis:2005zk,Drepanou:2021jiv},
\begin{align}
\label{kani}
\mathcal{S}_\mathrm{K} = \frac{1}{K} \sinh \left(K \mathcal{S} \right)\, ,
\end{align}
and it reduces to the Bekenstein-Hawking entropy as the parameter $K \to 
0$.

 Completely different motivations led to the introduction of the Barrow 
entropy, which was designed \cite{Barrow:2020tzx} to 
describe spacetime foam in quantum gravity,
\begin{align}
\label{barrow}
\mathcal{S}_\mathrm{B} = \left( 
\frac{A}{A_\mathrm{Pl}}\right)^{1+ \, \Delta/2}\,.
\end{align}
The event horizon has area $A$, while  $A_\mathrm{Pl}\equiv 4G$ 
is the Planck 
area, and the  exponent $\Delta$ embodies the quantum gravity 
deformation. Maximal quantum deformation corresponds to  
$\Delta = 1$, while $ \mathcal{S}_\mathrm{B}$ becomes the 
usual Bekenstein-Hawking entropy if $\Delta \rightarrow 0$,

The last entropy that we consider, different from the previous ones, 
originated in non-extensive statistical 
mechanics applied to Loop Quantum Gravity 
\cite{Majhi:2017zao,Czinner:2015eyk,Mejrhit:2019oyi, Liu:2021dvj}. It is
\begin{align}
\mathcal{S}_q= \frac{1}{1-q} \left[ \mathrm{e}^{ (1-q)\Lambda(\gamma_0) 
\mathcal{S} } -1 \right] \,.\label{LQGentropy}
\end{align}
where  $q$ weights the 
probability of very frequent events differently than that of infrequent 
ones. 
Moreover, 
\begin{align}
\Lambda( \gamma_0) = \frac{ \ln 2}{\sqrt{3} \, \pi \gamma_0}  
\end{align}
and $\gamma_0$ is the Barbero-Immirzi parameter. The use of  
different gauge groups attributes one of the two values $ 
\frac{\ln2}{\pi 
\sqrt{3}}$ or $\frac{\ln 
3}{2\pi \sqrt{2} }$ to $\gamma_0$. In scale-invariant 
gravity $\gamma_0 $ becomes a free parameter \cite{Veraguth:2017uwp, 
Wang:2018bdg, Wang:2019ryx}. The 
choice $\gamma_0= \frac{\ln2}{\pi \sqrt{3}} $ gives  
$\Lambda(\gamma_0)=1$ and makes the Loop Quantum Gravity 
entropy~(\ref{LQGentropy}) reproduce extensive statistical 
mechanics. $\mathcal{S}_q$ becomes the Bekenstein-Hawking entropy if 
$q= 1$. The entropy~(\ref{LQGentropy}) was used for black holes in 
\cite{Majhi:2017zao, Czinner:2015eyk, Mejrhit:2019oyi} and in cosmology in 
\cite{Liu:2021dvj}.

Two more generalizations of entropy were proposed recently in 
\cite{Nojiri:2022aof}. The first is the six-parameter entropy 
\begin{align}
\label{general1}
\mathcal{S}_\mathrm{G} \left( \alpha_\pm, \beta_\pm, \gamma_\pm \right)
= \frac{1}{\alpha_+ + \alpha_-}
\left[ \left( 1 + \frac{\alpha_+}{\beta_+} \, \mathcal{S}^{\gamma_+} 
\right)^{\beta_+} - \left( 1 + \frac{\alpha_-}{\beta_-} 
\, \mathcal{S}^{\gamma_-} \right)^{-\beta_-} \right] \,,
\end{align}
where all the parameters $\left( \alpha_{\pm} , 
\beta_{\pm}, \gamma_{\pm} \right) $ are positive. For suitable values 
of the parameters, this entropy 
reduces to the entropies~(\ref{TS1}), (\ref{RS1}), (\ref{SM}), 
(\ref{barrow}), (\ref{kani}), and (\ref{LQGentropy}) previously reported. 
Taking  $\alpha_+=\alpha_- = 0$ and  
$\gamma_{-}= \gamma_{+} \equiv \gamma$, the  values 
$\gamma=\delta$ or $\gamma= 1 + \Delta/2 $ reproduce the Tsallis 
entropy~(\ref{TS1}) and the Barrow entropy (\ref{barrow}).  If 
$\alpha_-=0$ and we write 
$\alpha_+=R$, $\beta_+ = R/\delta$, and 
$\gamma_+=\delta$, the Sharma-Mittal entropy~(\ref{SM}) is obtained. 
 If instead one takes the limit $\alpha_+ \rightarrow 0$ and
$\beta_+ \rightarrow 0$ with $\alpha \equiv \alpha_+ / \beta_+ $ 
finite, then by setting $\gamma_+=1$ one recovers the R{\'e}nyi 
entropy~(\ref{RS1}). The other limit $\beta_\pm\to 0$ of the general 
entropy~(\ref{general1}) with $\gamma_\pm=1$ and $\alpha_\pm = K$ 
reduces it to the Kaniadakis entropy~(\ref{kani}). 
Finally,  setting $\alpha_{-}=0 $ and 
$\gamma_{+}=1$ in~(\ref{general1}), the limit 
$\beta_{+} \rightarrow +\infty $ in conjunction with $\alpha=1-q$ gives 
back the Loop Quantum Gravity entropy~(\ref{LQGentropy}) with $\Lambda( 
\gamma_0 ) =1$. 

%The latter becomes the Bekenstein-Hawking entropy 
%$\mathcal{S}$ if $q= 1$.

The second proposal advanced in \cite{Nojiri:2022aof} contains only 
three parameters: 
\begin{align}
\label{general6}
\mathcal{S}_\mathrm{G} \left( \alpha, \beta, \gamma \right)
= \gamma^{-1} \left[ \left( \frac{\alpha}{\beta} \, \mathcal{S} +1  
\right)^\beta - 1 \right] \,.
\end{align}
Again, $ \alpha , \beta$, and $ \gamma $ are positive. When 
$\gamma=\alpha$, $\mathcal{S}_\mathrm{G}$ coincides with the 
Sharma-Mittal entropy~(\ref{SM}) with 
$\mathcal{S}_\mathrm{T}=\mathcal{S}$ and $\delta=1$. If we set $\gamma = 
\left( \alpha/\beta \right)^{\beta} $, then~(\ref{general6}) 
reduces to the Tsallis entropy~(\ref{TS1}) if $\beta=\delta$ and to the 
Barrow entropy~(\ref{barrow}) if $\alpha\to \infty$. Finally, the limit 
$\left( \alpha, \beta \right) \rightarrow \left( 0, 0 \right) $  with $ 
\alpha/\beta $ 
finite yields the R{\'e}nyi entropy~(\ref{RS1}), provided that 
$\alpha/\beta$ is replaced by $\alpha$ and that $\gamma=\alpha$.

If spherical spacetimes are considered in conjunction with the 
Tsallis entropy (\ref{TS1}), Eq.~(\ref{entropy5}) becomes
\begin{align}
\label{entropy6}
\frac{h_2\left(r=r_\mathrm{h}; c_i\left(r_\mathrm{h}\right)\right) } 
{\left(1 - \left. \frac{\partial m\left(r; c_i\left(r_\mathrm{h}\right) 
\right)}{\partial r}\right|_{r= r_\mathrm{h}}\right)^2}
= \delta^2 \left( \frac{4\pi {r_\mathrm{h}}^2}{A_0} 
\right)^{2\left(\delta - 1 \right)} 
\end{align}
while, for the same geometry, the R{\'e}nyi 
entropy~(\ref{RS1}) gives 
\begin{align}
\label{entropy8}
\frac{h_2\left(r=r_\mathrm{h}; c_i\left(r_\mathrm{h}\right)\right) } 
{\left(1 - \left. \frac{\partial m\left(r; c_i\left(r_\mathrm{h}\right) 
\right)}{\partial r}\right|_{r= r_\mathrm{h}}\right)^2}
= \frac{1}{ \left( 1 + \frac{\pi \alpha {r_\mathrm{h}}^2}{G} \right)^2} 
\, .
\end{align}
The Kaniadakis entropy~(\ref{kani}) leads to 
\begin{align}
\label{kani2}
\frac{h_2\left(r=r_\mathrm{h}; c_i\left(r_\mathrm{h}\right)\right) } 
{\left(1 - \left. \frac{\partial m\left(r; c_i\left(r_\mathrm{h}\right) 
\right)}{\partial r}\right|_{r= r_\mathrm{h}}\right)^2}
= \cosh^2 \left(\frac{\pi K {r_\mathrm{h}}^2}{G} \right) 
\end{align}
and the six-parameter entropy  
(\ref{general1}) yields
\begin{align}
\label{general1h2}
\frac{h_2\left(r=r_\mathrm{h}; c_i\left(r_\mathrm{h}\right)\right) } 
{\left(1 - \left. \frac{\partial m\left(r; c_i\left(r_\mathrm{h}\right) 
\right)}{\partial r}\right|_{r= r_\mathrm{h}}\right)^2} 
=\, & \frac{1}{\left(\alpha_+ + \alpha_-\right)^2}
\left[ \alpha_+ \gamma_+ \left(\frac{\pi {r_\mathrm{h}}^2}{G} 
\right)^{\gamma_+ -1}\left( 1 + \frac{\alpha_+}{\beta_+} 
\left(\frac{\pi {r_\mathrm{h}}^2}{G} \right)^{\gamma_+} \right)^{\beta_+ 
- 1} \right. \nonumber \\
&\, \left. + \alpha_- \gamma_- \left(\frac{\pi {r_\mathrm{h}}^2}{G} 
\right)^{\gamma_- -1}\left( 1 + \frac{\alpha_-}{\beta_-} 
\left(\frac{\pi {r_\mathrm{h}}^2}{G} \right)^{\gamma_-} \right)^{-\beta_- 
- 1} \right]^2 \,.
\end{align}
The simplified three-parameter entropy~(\ref{general6}) gives 
\begin{align}
\label{general5h2}
\frac{h_2\left(r=r_\mathrm{h}; c_i\left(r_\mathrm{h}\right)\right) } 
{\left(1 - \left. \frac{\partial m\left(r; c_i\left(r_\mathrm{h}\right) 
\right)}{\partial r}\right|_{r= r_\mathrm{h}}\right)^2}
= \frac{\alpha^2}{\gamma^2} \left[ 1 + \left(\frac{\pi\alpha 
{r_\mathrm{h}}^2}{\beta G} \right) \right]^{2\beta - 2} \,.
\end{align}

These possibilities may seem rather abstract, therefore in the next 
section, we provide concrete models that realize these relations.

\section{Spherically symmetric solutions of Einstein-two-scalar models} 
\label{sec:4}

Our first model consists of general relativity with a scalar 
doublet $ \left( \phi, \chi \right)$ as the matter source and with action
\begin{align} 
\label{I8} S_{( \mathrm{GR} \phi\chi)} = \int d^4 x \sqrt{-g} & \left[ 
\frac{R}{2\kappa^2}
 - \frac{ A (\phi,\chi)}{2} \, \partial_{\alpha} \phi \, \partial^{\alpha} 
\phi - B (\phi,\chi) \, \partial_{\alpha} \phi \, \partial^{\alpha} \chi 
\right. \nn 
& \left. \quad - \frac{ C (\phi,\chi)}{2} \, \partial^{\alpha} \chi 
\partial_{\alpha} \chi - \, V \left( \phi,\chi \right) \right] \, , 
\end{align} 
where $V(\phi, \chi)$ is the two scalars interaction potential, while  
$A$, $B$, and $C$ depend on both of them.  The energy-momentum tensor is 
\begin{align} 
\label{I9} 
T^{(\phi\chi)}_{\mu\nu} =& g_{\mu\nu} \left[
 - \frac{ A \left( \phi,\chi \right)}{2} \, \partial_\rho \phi \, 
\partial^\rho \phi
 - \, B \left(\phi,\chi \right) \, \partial_\rho \phi \, \partial^\rho 
\chi
 - \frac{ C (\phi,\chi)}{2} \, \partial^{\alpha} \chi \, \partial_{\alpha}  
\chi - V(\phi,\chi)\right] \nn
& + A (\phi,\chi) \, \partial_\mu \phi \, \partial_\nu \phi
+ B (\phi,\chi) \left( \partial_\mu \phi \, \partial_\nu \chi
+ \partial_\nu \phi \, \partial_\mu \chi \right)
+ C (\phi,\chi) \, \partial_\mu \chi \, \partial_\nu \chi 
\end{align} 
and there are two conservation equations stemming from  the contracted 
Bianchi identities, 
\begin{eqnarray}
\label{I10} 
& \frac{A_\phi}{2} \, \partial_\mu \phi \, \partial^\mu \phi
+ A \nabla^\mu \partial_\mu \phi + A_\chi \, \partial_\mu \phi \, 
\partial^\mu \chi 
+ \left( B_\chi - \frac{1}{2} \, C_\phi \right)\partial_\mu \chi \, 
\partial^\mu \chi \nonumber\\
&+ B \nabla^\mu \partial_\mu \chi - V_\phi  =0 \,,\\
\label{I10b} 
& \left( - \frac{A_\chi}{2} + B_\phi \right) 
\partial_\mu \phi \, \partial^\mu \phi + B \nabla^\mu \partial_\mu \phi
+ \frac{C_\chi}{2} \, \partial_\mu \chi \, \partial^\mu \chi + C 
\nabla^\mu \partial_\mu \chi \nonumber\\
&\, + C_\phi \, \partial_\mu \phi \, \partial^\mu \chi - V_\chi=0 \, , 
\end{eqnarray} 
where $A_\phi \equiv \partial A(\phi,\chi)/\partial \phi$. 
We use the identifications 
\begin{align} 
\label{TSBH1} \phi= \, t\, , \quad\quad \chi \, =r 
\end{align} 
without loss of generality because, for general spherical  
solutions, $\phi$ and $\chi$ depend on  $t$ and $r$ and spherical 
solutions of this model contains  $\phi(t,r), 
\chi(t,r)$ explicitly.  One can then invert these 
functions in spacetime regions in which they are 
one-to-one, $\partial_{\mu} \phi $ is timelike, and 
$\partial_{\mu} \chi$ is spacelike. Then the scalar fields are redefined, 
trading $\phi$ and $\chi$ with  $t$ and $r$. The latter now play the role 
of redefined scalar  fields $\bar{\phi}$ 
and $\bar{\chi}$ (conversely, $\phi(t,r)\to 
\phi(\bar{\phi}, \bar{\chi})$ 
and $\chi(t,r) \to \chi(\bar{\phi}, \bar{\chi})$). 
These new fields can be identified with $t$ and $r$ (see 
Eq.~(\ref{TSBH1})). 
The  variables change $\left( \phi , \chi\right) \rightarrow \left( 
\bar{\phi} , \bar{\chi} \right)$ is now incorporated into redefinitions 
of the coefficients $A$, $B$, $C$, and $V$ in the action 
integral~(\ref{I8}). Thus, under mild assumptions, the {\em 
ansatz}~(\ref{TSBH1}) does not imply loss of generality.

Continue the analysis of the Einstein equations sourced by this scalar 
doublet: their time-time, radius-radius,  
$\left(i,j\right)$, and time-radius  components read (the remaining 
equations are trivially satisfied because of the 
spherical symmetry) 
\begin{align} 
\label{TSBH2}
& \frac{\e^{2\left(\nu - \lambda\right)}}{\kappa^2}
\left( \frac{2\lambda'}{r} + \frac{\e^{2\lambda} - 1}{r^2} \right) 
= - \e^{2\nu} \left( - \frac{A}{2} \, \e^{-2\nu} - \frac{C}{2} \, 
\e^{-2\lambda} - V \right) \, ,\\
\label{TSBH3}
& \frac{1}{\kappa^2} \left( \frac{2\nu'}{r} - \frac{\e^{2\lambda}
 - 1}{r^2} \right) = \e^{2\lambda} \left( \frac{A}{2} \, \e^{-2\nu}
+ \frac{C}{2} \, \e^{-2\lambda} - V \right) \, ,\\ 
\label{TSBH4}
& \frac{1}{\kappa^2} \left[ - \e^{-2 \nu} \left\{ \ddot\lambda
+ \left( \dot\lambda - \dot\nu \right) \dot\lambda \right\}
+ \e^{-2\lambda}\left( r \left(\nu' - \lambda' \right)
+ r^2 \nu'' + r^2 \left( \nu' - \lambda' \right) \nu' \right) \right] \nn
& \qquad = r^2 \left( \frac{A}{2} \, \e^{-2\nu} - \frac{C}{2} \,
\e^{-2\lambda} - V \right) \, ,\\
\label{TSBH5}
& \dot\lambda = \frac{\kappa^2 r B}{2}  \,.
\end{align} 
The solutions of  Eqs.~(\ref{TSBH2})-(\ref{TSBH5}) are written as 
\begin{align} 
\label{TSBH6} 
A(t,r)=& \,\frac{1}{\kappa^2} \left[ - \left\{ \ddot\lambda
+ \dot\lambda  \left( \dot\lambda - \dot\nu \right)  \right\} 
\right.\nonumber\\
& \left.
+  \e^{2\left(\nu - \lambda\right)} \left( \frac{\e^{2\lambda} - 1}{r^2}
+  \frac{\lambda' + \nu' }{r} + \nu'' + \nu' \left( \nu' - \lambda' 
\right)  
\right) \right] \, , \\
\label{TSBH7} 
B(t,r)=& \, \frac{2\dot\lambda}{\kappa^2 r} \, , \\
\label{TSBH8} 
C(t,r)=& \, \frac{1}{\kappa^2} \left[ \frac{\e^{-2\left(\nu - \lambda 
\right)}}{r^2}
\left\{ \ddot\lambda + \left( \dot\lambda - \dot\nu \right) \dot\lambda 
\right\} - \frac{\e^{2\lambda} - 1}{r^2}
+ \frac{\nu' + \lambda'}{r} - \nu'' \right.\nonumber\\
&\left.  + \nu' \left(  \lambda' -\nu' \right)  \right] \, , \\
\label{TSBH9} 
V(t,r)=& \, \frac{\e^{-2\lambda}}{2\kappa^2} \left[ \frac{2\left( \lambda' - 
\nu'\right) }{r} + \frac{2 \left(\e^{2\lambda} - 1\right) }{r^2} \right] 
\,.
\end{align} 
$A$,  $B$, $C$, $V$ depend on $\left( \phi , \chi \right)$
because they depend on $t$ and $r$ that, in turn, depend on 
$\left( \phi , \chi \right) $.  
Conversely, assigning $A$, $B$, $C$, $V$ the model admits 
spherical solutions~(\ref{metric}) such that  
the functions $\nu$ and $\lambda$ are arbitrary.

For {\em static} spacetime geometries, $A$, $B$, $C$, 
and  $V$ are time-independent or, equivalently, they only depend 
on $\phi$ and  Eqs.~(\ref{TSBH6})-(\ref{TSBH9}) reduce to
\begin{align}
\label{TSBH6s} 
A=& \,\frac{1}{\kappa^2} \e^{2\left(\nu - \lambda\right)} \left[  
\frac{\e^{2\lambda} - 1}{r^2} + \frac{ \lambda'+ \nu' }{r}  - \nu' \left(  
\lambda'-\nu' \right)   + \nu''  \right] \, , \\
\label{TSBH7s} 
B=& \, 0 \, , \\
\label{TSBH8s} 
C=& \, \frac{1}{\kappa^2} \left[ \frac{1- \e^{2\lambda} }{r^2}
+ \frac{  \lambda' +\nu' }{r}  + \nu' \left(\lambda' - \nu'  \right)  
- \nu'' \right] \, , \\
\label{TSBH9s} 
V=& \, \frac{\e^{-2\lambda}}{\kappa^2} \left[ \frac{\e^{2\lambda} - 
1}{r^2} - \frac{\nu' - \lambda'}{r} \right] \, . 
\end{align} 
Assuming 
\begin{align}
\label{exam1}
\e^{2\nu}= h= \frac{1}{h_2(r)} \left( 1 - \frac{r_\mathrm{h}}{r} \right) 
\, , \quad \e^{-2\lambda}= h_1 = h_2 h = 1 - \frac{r_\mathrm{h}}{r} 
\end{align}
and replacing $r$ with $\phi$, the functions $A$, $B$, $C$, 
and $V$ become
\begin{align}
\label{TSBH6sB} 
A(\phi)=& \,\frac{1}{\kappa^2{h_2(\phi)}^2} \left( 1 - 
\frac{r_\mathrm{h}}{\phi} \right)^2 
\left[ - \frac{h_2'(\phi)}{4\phi h_2(\phi)} + 
\frac{3h_2' (\phi)}{4 h_2(\phi) \left(\phi - r_\mathrm{h}\right) } 
+ \frac{h_2'' (\phi)}{2h_2(\phi)} \right.\nonumber\\
&\left.   - 
 \left(\frac{h_2'(\phi)}{ 2h_2(\phi)}\right)^2 \right] \, , \\
\label{TSBH7sB} 
B(\phi)=& \, 0 \, , \\
\label{TSBH8sB} 
C(\phi)=& \,\frac{1}{ \kappa^2} \left[ \frac{5h_2'(\phi)}{4\phi h_2(\phi)} 
- 
\frac{3h_2'(\phi)}{4\left(\phi - r_\mathrm{h}\right) h_2(\phi)} 
 - \frac{h_2'' (\phi)}{2h_2(\phi)} + 
\frac{1}{4}\left( \frac{h_2'(\phi)}{h_2(\phi)}\right)^2 \right] \, , \\
\label{TSBH9sB} 
V(\phi)=& \, \frac{1}{2\kappa^2 \phi} \left( 1 - 
\frac{r_\mathrm{h}}{\phi} 
\right) \frac{h_2'(\phi)}{h_2(\phi)} \, . 
\end{align} 
It is important that $A$, $C$, and $V$ depend explicitly on 
the horizon radius  $r_\mathrm{h}$, hence the latter 
 is fixed in this model. There should be other solutions in 
addition to those corresponding to Eq.~(\ref{exam1}), but they may be 
difficult to find explicitly. This problem is bypassed using the trick of 
Ref.~\cite{Nojiri:2017kex} of adding to the Lagrangian the term
$
\mathcal{L}_{\rho\sigma}=\rho^\mu \partial_\mu \sigma $. Varying 
$\mathcal{L}_{\rho\sigma}$ with respect to $\rho^\mu$ 
yields constant $\sigma$,
\begin{align}
\label{rhosigma2}
 \partial_\mu \sigma =0 \,,
\end{align}
We can identify $\sigma $ with 
the horizon radius $ r_\mathrm{h} $. Then, replacing $r_\mathrm{h}$ 
with $\sigma $ in the functions $A$, $C$, 
and $V$ of Eqs.~(\ref{TSBH6sB}), (\ref{TSBH8sB}), and (\ref{TSBH9sB}), 
$r_\mathrm{h}$ plays the role of an integration constant appearing in 
Eq.~(\ref{rhosigma2}): 
\begin{align}
\label{TSBH6sC} 
A(\phi,\sigma)=& \,\frac{1}{\kappa^2{ {h_2}^2(\phi,\sigma)} } \left(  
\frac{\sigma}{\phi} -1 \right)^2 
\left[ - \frac{ h_{2,\phi} }{4\phi \, h_2(\phi,\sigma)} + 
\frac{3 \, h_{2,\phi} }{4\left(\phi - \sigma\right) 
h_2(\phi,\sigma)} \right. \nonumber\\
& \left. 
+ \frac{h_{2,\phi\phi} }{2 \, h_2(\phi,\sigma)} - 
\frac{1}{4}\left(\frac{h_{2,\phi} }{h_2(\phi,\sigma)}\right)^2 
\right] \, , \\
\label{TSBH7sC} 
B(\phi,\sigma)=& \, 0 \, , \\
\label{TSBH8sC} 
C(\phi,\sigma)=& \,\frac{1}{\kappa^2} \left[ 
\frac{5h_2(\phi,\sigma)_{,\phi}}{4\phi h_2(\phi,\sigma)} - 
\frac{3h_2(\phi,\sigma)_{,\phi}}{4\left(\phi - \sigma\right) 
h_2(\phi,\sigma)} 
 - \frac{h_2(\phi,\sigma)_{,\phi\phi}}{2h_2(\phi,\sigma)} + 
\left(\frac{h_2(\phi,\sigma)_{,\phi}}{2h_2(\phi,\sigma)} 
\right)^2 \right] , \\
\label{TSBH9sC} 
V(\phi,\sigma)=& \, \frac{1}{2\kappa^2 \phi} \left( 1 - 
\frac{\sigma}{\phi} \right) 
\frac{h_2(\phi,\sigma)_{,\phi}}{h_2(\phi,\sigma)} \,,
\end{align}
where $ 
h_2(\phi,\sigma)_{,\phi} \equiv \partial 
h_2(\phi,\sigma) / \partial\phi $, $ h_2(\phi,\sigma)_{,\phi\phi} \equiv 
\partial^2 h_2(\phi,\sigma) / \partial\phi^2$. 

While, in Einstein gravity, the Kerr geometry is the unique vacuum and 
asymptotically flat solution, in the present model the matter scalar 
doublet acts as hair. Hairy black holes  are common in alternative  
theories of  gravity, for example ``first generation'' scalar-tensor  and 
more modern Horndeski theories.\footnote{The Reissner-Nordstr{\"o}m black 
hole  already differs from the Schwarzschild one due to the Maxwell field 
sourcing the Eintein 
equations.}

Under the assumption~(\ref{exam1}), the effective mass(\ref{MDmassfunc})
becomes  the constant $m_\mathrm{eff}(r)=r_\mathrm{h}/(2G)$, 
(\ref{entropy5}) reduces to 
\begin{align}
\label{entropy5C1}
h_2\left(r=r_\mathrm{h}; c_i\left(r_\mathrm{h}\right)\right) = 16G^2 
\Big[ \mathcal{S}_\mathrm{bh}' \left(A\right) \Big]^2 
\end{align} 
and choosing, as an example, 
\begin{align}
\label{entropy5C2}
h_2(\phi,\sigma)= 1 - \frac{\sigma}{\phi}\left[ 1 - 16G^2 \left( 
\mathcal{S}_\mathrm{bh}' \left( 4\pi \sigma^2 \right) \right)^2 \right] 
\,,
\end{align}
we obtain
\begin{align}
\label{entropy5C3}
h_2\left(\phi=r_\mathrm{h}, \sigma=r_\mathrm{h}\right) = 
16G^2 \Big[ \mathcal{S}_\mathrm{bh}' \left(A\right) \Big]^2 
\end{align}
as in (\ref{entropy5C1}). The condition 
\begin{align}
\label{entropy5C4}
h_2\left(\phi=r\to \infty, \sigma=r_\mathrm{h}\right) = 1 
\end{align}
is required by asymptotic flatness. As a conclusion, several of the 
generalized entropies in the last section can be realized with the 
choice~(\ref{entropy5C2}).

For the Tsallis entropy (\ref{TS1}), we find 
\begin{align}
\label{entropy6B}
h_2(\phi,\sigma)= 1 - \frac{\sigma}{\phi}\left[ 1 - \delta^2 \left( 
\frac{4\pi {\sigma}^2}{A_0} \right)^{2\left(\delta - 1 \right)} \right] 
\end{align}
while, to include other examples, the R{\'e}nyi 
entropy~(\ref{RS1}) gives
\begin{align}
\label{entropy8B}
h_2(\phi,\sigma)= 1 - \frac{\sigma}{\phi}\left[ 1 - \frac{1}{ \left( 1 + 
\frac{\pi \alpha {\sigma}^2}{G} \right)^2} \right] \,,
\end{align}
and the Kaniadakis entropy~(\ref{kani}) leads to 
\begin{align}
\label{kani3}
h_2(\phi,\sigma)= 1 - \frac{\sigma}{\phi}\left[ 1 - \cosh^2 
\left(\frac{\pi K {\sigma}^2}{G} \right) \right] \,.
\end{align}
The six-parameter entropy~(\ref{general1}) and the three-parameter entropy 
(\ref{general6}) give, respectively, 
\begin{align}
\label{general1h2B}
h_2(\phi,\sigma)= 1 \nonumber\\
- \frac{\sigma}{\phi} &\, \left\{ 1 - 
\frac{1}{\left(\alpha_+ + \alpha_-\right)^2}
\left[ \alpha_+ \gamma_+ \left(\frac{\pi {\sigma }^2}{G} 
\right)^{\gamma_+ -1}\left( 1 + \frac{\alpha_+}{\beta_+} 
\left(\frac{\pi {\sigma }^2}{G} \right)^{\gamma_+} \right)^{\beta_+ - 1} 
\right. \right. \nonumber \\
& \left. \left. + \alpha_- \gamma_- \left(\frac{\pi {\sigma }^2}{G} 
\right)^{\gamma_- -1}\left( 1 + \frac{\alpha_-}{\beta_-} 
\left(\frac{\pi {\sigma }^2}{G} \right)^{\gamma_-} \right)^{-\beta_- - 1} 
\right]^2 \right\} \,,
\end{align}
and
\begin{align}
\label{general5h2}
h_2(\phi,\sigma)= 1 - \frac{\sigma}{\phi}\left[ 1 - 
\frac{\alpha^2}{\gamma^2} \left( 1 + \left(\frac{\pi\alpha {\sigma 
}^2}{\beta G} \right) \right)^{2\beta - 2} \right] \,. 
\end{align}

%Hence models can be constructed with the above-mentioned entropies 
%alternative to the Bekenstein-Hawking entropy.

\subsection{Thermodynamics of the model (\ref{general5h2})}

Let us examine the thermodynamics of the model~(\ref{general5h2}) and 
compare it with the one obtained with the Bekenstein-Hawking 
entropy~(\ref{S3}) for the solution~(\ref{exam1}), (\ref{general5h2}).

For the solution $\phi=r$ and $\sigma=r_\mathrm{h}$ Eq.~(\ref{entropy3d}) 
gives, at the horizon $r=r_\mathrm{h}$, 
\begin{align}
\label{entropy3d2}
\mathcal{S}_\mathrm{bh} = \frac{1}{2G}\int_0^{r_\mathrm{h}} d\xi 
\, 4\pi \xi \left\{ 
\frac{\alpha}{\gamma} \left[ 1 + \left(\frac{\pi\alpha \xi^2}{\beta G} 
\right) \right]^{\beta - 1}\right\} \, .
\end{align}
Now $ m(r)=M= r_\mathrm{h}/(2G)$ is constant and, because the 
thermodynamical energy is $E=M= r_\mathrm{h} /( 2G)$, it is  
\begin{align}
\label{Tex3}
\frac{1}{T}=\frac{d \mathcal{S}_\mathrm{bh} }{dE}= \left\{ 
\frac{\alpha}{\gamma} \left[ 1 + \left(\frac{\pi\alpha 
r_\mathrm{h}^2}{\beta G} \right) \right]^{\beta - 1}\right\} 4\pi 
r_\mathrm{h} \,,
\end{align}

\begin{align}
\label{TH6B}
T = \frac{C\left( r_\mathrm{h} \right) T_\mathrm{H}}{\sqrt{h_2 \left( 
r_\mathrm{h} \right)}} 
= \frac{ \gamma \, T_\mathrm{H}}{ \alpha \left[ 1 + 
\left(\frac{\pi\alpha r_\mathrm{h}^2}{\beta G} \right) \right]^{\beta - 
1}}\,,
\end{align} 
where the last equation follows from Eq.~(\ref{TH6}). Here $C\left( 
r_\mathrm{h} \right)=1$ because $m(r)=M$ is constant and, since 
\begin{align}
\label{HTB1}
T_\mathrm{H}\equiv \frac{1}{8\pi G m_\mathrm{eff} \left( r_\mathrm{h} 
\right)}
= \frac{1}{8\pi G m \left( r_\mathrm{h} \right)} = \frac{1}{8\pi G M} = 
\frac{1}{4\pi r_\mathrm{h}}\,,
\end{align}
the expressions~(\ref{Tex3}) and~(\ref{TH6B}) coincide. The 
integral~(\ref{entropy3d2}) is computed explicitly giving the 
three-parameter entropy~(\ref{general6}) as
\begin{align}
\label{entropy3d3}
\mathcal{S}_\mathrm{bh} = \gamma^{-1} \left\{ \left[ 1 + 
\left(\frac{\pi\alpha {r_\mathrm{h}}^2}{\beta G} \right) \right]^\beta -1 
\right\}
= \gamma^{-1} \left[ \left( 1 + \frac{\alpha}{\beta} \mathcal{S} 
\right)^\beta - 1 \right] \,.
\end{align}
 
When we use the Bekenstein-Hawking entropy~(\ref{S3}), we need to 
specify the energy or the temperature. Postulating that   
the internal energy $E$ is the mass function~(\ref{MDmassfunc}), 
for the solution~(\ref{exam1}), (\ref{general5h2}) one finds $E=M$ 
because $m_\mathrm{eff}(r)=M=$const. and the thermodynamic 
relation $\frac{1}{T}=\frac{d \mathcal{S}}{dE}$ gives the standard 
Hawking temperature $T_\mathrm{H}$, which is different from the 
temperature~(\ref{TH6B}). If, instead, (\ref{TH6B}) is adopted as the 
temperature, 
\begin{align}
\label{TH6C}
T = \frac{T_\mathrm{H}}{\frac{\alpha}{\gamma} \left[ 1 + 
\left(\frac{\pi\alpha r_\mathrm{h}^2}{\beta G} \right) \right]^{\beta - 
1}}
= \frac{\gamma}{ 4\pi \alpha r_\mathrm{h} \left[ 1 + 
\left(\frac{\pi\alpha r_\mathrm{h}^2}{\beta G} \right) \right]^{\beta - 
1}} \, ,
\end{align}
then 
\begin{align}
\label{dE}
dE=\frac{d\mathcal{S}}{T}= \frac{4\pi^2 \alpha {r_\mathrm{h}}^3}{\gamma 
G} \left[ 1 + \left(\frac{\pi\alpha r_\mathrm{h}^2}{\beta G} \right) 
\right]^{\beta - 1} dr_\mathrm{h} \nonumber\\
= \frac{4\pi \beta}{\gamma} \left\{ \left[ 1 + \left(\frac{\pi\alpha 
r_\mathrm{h}^2}{\beta G} \right) \right)^\beta 
 - \left( 1 + \left(\frac{\pi\alpha r_\mathrm{h}^2}{\beta G} \right) 
\right]^{\beta - 1} \right\} r_\mathrm{h} dr_\mathrm{h} 
\end{align}
and the internal energy is 
\begin{align}
\label{E} 
E = \frac{2 \beta^2 G}{\alpha\gamma} 
\left\{ \frac{1}{\beta + 1} \left[ 1 + \left(\frac{\pi\alpha 
r_\mathrm{h}^2}{\beta G} \right) \right]^{\beta + 1} 
- \frac{1}{\beta} \left[ 1 + \left(\frac{\pi\alpha r_\mathrm{h}^2}{\beta 
G} \right) \right]^\beta \right\} + E_0 \,,
\end{align}
where $E_0$ is an integration constant. The condition $E ( 
r_\mathrm{h}=0)=0 $ determines this integration constant,   
\begin{align}
\label{E0}
E_0 = - \frac{2 \beta^2 G}{\alpha\gamma} 
\left( \frac{1}{\beta + 1} - \frac{1}{\beta} \right) \, .
\end{align}
In any case, the expression of $E$ obtained is not 
$M= r_\mathrm{h} /(2G)$ as in Eq.~(\ref{entropy5C1}).

\section{Conclusions}
\label{sec:5}

When applied to black holes, entropy notions alternative to the 
Bekenstein-Hawking entropy  usually 
lead to inconsistencies in the thermodynamics because the 
relation $TdS=dE $ linking entropy, energy, and temperature is violated  
if temperature and internal energy are the Hawking 
temperature and the black hole mass. However, consistency can be achieved 
for non-Schwarzschild black holes in modified gravity if  
the horizon radius (consequently, the size of its area appearing in 
Bekenstein's area law), are modified. This is precisely what we have 
explored in the previous sections. Indeed, consistency of new entropy 
proposals with Hawking temperature and area law is possible for certain 
analytical black hole solutions, reported above. 

While it is interesting that the consistency problem of black hole 
thermodynamics can actually be cured in this way, the examples shown here 
are not a panacea and there are certain limitations. We have restricted 
the scope of this investigation to spherical, static, and 
asymptotically flat black holes of modified gravity. Strictly 
speaking, modified gravity theories that are interesting to model the 
current expansion of the universe without dark  
energy have a built-in, time-varying, cosmological ``constant'' and are 
not asymptotically flat. However, it is true that (possibly with the 
exception of some primordial black holes), the cosmological dynamics can 
be safely neglected near the black hole horizon, the spacetime region 
where we set our discussion. Therefore, on physical grounds, this is not a 
fundamental limitation.

The exploration of specific theories of gravity, for example Horndeski or 
Degenerate Higher Order (DHOST) theories involves plenty of details and 
this is the reason why we have not committed to any specific theory here, 
but we have provided a general, although perhaps preliminary, discussion. 
We have focused on the possibility of changing the black hole mass, but 
the strength of the gravitational coupling $G_\mathrm{eff}$ varies as well 
in many modified gravities and can be used to one's advantage to search 
for viable thermodynamics with generalized entropy, area law, black hole 
radii deviating from the Schwarzschild radius, and non-standard 
gravitational couplings (this is not an option, for example, in the 
Einstein-scalar doublet model of Sec.~\ref{sec:4}). For example, in 
Horndeski gravity the effective gravitational coupling becomes a function 
of both the gravitational scalar $\phi$ of the theory and of 
$\nabla^{\mu}\phi \nabla_{\mu} \phi$. Specific theories, their known black 
hole solutions, and the possibility of consistent black hole 
thermodynamics will be analized elsewhere.

\section*{Acknowledgments}

This work is partially supported by JSPS Grant-in-Aid for Scientific 
Research (C) No. 18K03615 (S.~N.), by MINECO (Spain) project 
PID2019-104397GB-I00 (S.~D.~O),  and by the Natural Sciences and 
Engineering Research Council of Canada grant~2016-03803 to V.~F.

%\appendix
%\section{Appendices}

%\section*{References}


\begin{thebibliography}{0}

%\cite{Bekenstein:1973ur}
\bibitem{Bekenstein:1973ur}
J.~D.~Bekenstein,
``Black holes and entropy,''
Phys. Rev. D \textbf{7} (1973), 2333-2346
doi:10.1103/PhysRevD.7.2333

%\cite{Hawking:1975vcx}
\bibitem{Hawking:1975vcx}
S.~W.~Hawking,
``Particle Creation by Black Holes,''
Commun. Math. Phys. \textbf{43} (1975), 199-220
[erratum: Commun. Math. Phys. \textbf{46} (1976), 206]
doi:10.1007/BF02345020

%\cite{Bardeen:1973gs}
\bibitem{Bardeen:1973gs}
J.~M.~Bardeen, B.~Carter and S.~W.~Hawking,
``The Four laws of black hole mechanics,''
Commun. Math. Phys. \textbf{31}, 161-170 (1973)
doi:10.1007/BF01645742

%\cite{Wald:1999vt}
\bibitem{Wald:1999vt} 
R.~M.~Wald,
``The thermodynamics of black holes,''
Living Rev. Rel. \textbf{4} (2001), 6 doi:10.12942/lrr-2001-6 
[arXiv:gr-qc/9912119 [gr-qc]].

%\cite{Carlip:2014pma}
\bibitem{Carlip:2014pma}
S.~Carlip,
``Black Hole Thermodynamics,''
Int. J. Mod. Phys. D \textbf{23}, 1430023 (2014)
doi:10.1142/S0218271814300237
[arXiv:1410.1486 [gr-qc]].

\bibitem{Wald} R.~M. Wald, {\em General Relativity} (Chicago University 
Press, Chicago, 1984).

%\cite{Nojiri:2021czz}
\bibitem{Nojiri:2021czz}
S.~Nojiri, S.~D.~Odintsov and V.~Faraoni,
``Area-law versus R\'enyi and Tsallis black hole entropies,''
Phys. Rev. D \textbf{104} (2021) no.8, 084030
doi:10.1103/PhysRevD.104.084030
[arXiv:2109.05315 [gr-qc]].
%6 citations counted in INSPIRE as of 14 Mar 2022

%%\cite{Tsallis:2012js}
%\bibitem{Tsallis:2012js}
%C.~Tsallis and L.~J.~L.~Cirto,
%%``Black hole thermodynamical entropy,''
%Eur. Phys. J. C \textbf{73} (2013), 2487
%doi:10.1140/epjc/s10052-013-2487-6
%[arXiv:1202.2154 [cond-mat.stat-mech]].
%%184 citations counted in INSPIRE as of 17 Aug 2021

\bibitem{Tsallis:1987eu}
C.~Tsallis, C, 
``Possible generalization of Boltzmann-Gibbs statistics''. 
Journal of Statistical Physics. 52 (1-2) (1988), 479-487 
doi:10.1007/BF01016429

%\cite{Ren:2020djc}
\bibitem{Ren:2020djc}
J.~Ren,
``Analytic critical points of charged R\'enyi entropies from hyperbolic 
%black holes,''
JHEP \textbf{05} (2021), 080
doi:10.1007/JHEP05(2021)080
[arXiv:2012.12892 [hep-th]].
%1 citations counted in INSPIRE as of 01 Aug 2021

%\cite{Nojiri:2019skr}
\bibitem{Nojiri:2019skr}
S.~Nojiri, S.~D.~Odintsov and E.~N.~Saridakis,
``Modified cosmology from extended entropy with varying exponent,''
Eur. Phys. J. C \textbf{79} (2019) no.3, 242
doi:10.1140/epjc/s10052-019-6740-5
[arXiv:1903.03098 [gr-qc]].
%50 citations counted in INSPIRE as of 17 Aug 2021

\bibitem{renyi}
A.~R{\'{e}}nyi, 
``On measures of information and entropy'' 
Proceedings of the Fourth Berkeley Symposium on Mathematics, Statistics 
and Probability, University of California Press (1960), 547-56

%\cite{Czinner:2015eyk}
\bibitem{Czinner:2015eyk}
V.~G.~Czinner and H.~Iguchi,
``R\'enyi Entropy and the Thermodynamic Stability of Black Holes,''
Phys. Lett. B \textbf{752} (2016), 306-310
doi:10.1016/j.physletb.2015.11.061
[arXiv:1511.06963 [gr-qc]].
%55 citations counted in INSPIRE as of 01 Aug 2021

%\cite{Tannukij:2020njz}
\bibitem{Tannukij:2020njz}
L.~Tannukij, P.~Wongjun, E.~Hirunsirisawat, T.~Deesuwan and C.~Promsiri,
``Thermodynamics and phase transition of spherically symmetric 
black hole in de Sitter space from R\'enyi statistics,''
Eur. Phys. J. Plus \textbf{135} (2020) no.6, 500
doi:10.1140/epjp/s13360-020-00517-2
[arXiv:2002.00377 [gr-qc]].
%3 citations counted in INSPIRE as of 01 Aug 2021

%\cite{Promsiri:2020jga}
\bibitem{Promsiri:2020jga}
C.~Promsiri, E.~Hirunsirisawat and W.~Liewrian,
``Thermodynamics and Van der Waals phase transition of charged 
black holes in flat space-time via R\'enyi statistics,''
Phys. Rev. D \textbf{102} (2020) no.6, 064014
doi:10.1103/PhysRevD.102.064014
[arXiv:2003.12986 [hep-th]].

%\cite{Samart:2020klx}
\bibitem{Samart:2020klx}
D.~Samart and P.~Channuie,
``AdS to dS phase transition mediated by thermalon in 
Einstein-Gauss-Bonnet gravity from R\'enyi statistics,''
[arXiv:2012.14828 [hep-th]].
%1 citations counted in INSPIRE as of 01 Aug 2021

%\cite{SayahianJahromi:2018irq}
\bibitem{SayahianJahromi:2018irq}
A.~Sayahian Jahromi, S.~A.~Moosavi, H.~Moradpour, J.~P.~Morais Gra\c{c}a, I.~P.~Lobo, 
I.~G.~Salako and A.~Jawad,
``Generalized entropy formalism and a new holographic dark energy model,''
Phys. Lett. B \textbf{780} (2018), 21-24
doi:10.1016/j.physletb.2018.02.052
[arXiv:1802.07722 [gr-qc]].
%105 citations counted in INSPIRE as of 19 Dec 2021

%\cite{Barrow:2020tzx}
\bibitem{Barrow:2020tzx}
J.~D.~Barrow,
``The Area of a Rough Black Hole,''
Phys. Lett. B \textbf{808} (2020), 135643
doi:10.1016/j.physletb.2020.135643
[arXiv:2004.09444 [gr-qc]].
%32 citations counted in INSPIRE as of 31 Jul 2021

%\cite{Kaniadakis:2005zk}
\bibitem{Kaniadakis:2005zk}
G.~Kaniadakis,
``Statistical mechanics in the context of special relativity. II.,''
Phys. Rev. E \textbf{72} (2005), 036108
doi:10.1103/PhysRevE.72.036108
[arXiv:cond-mat/0507311 [cond-mat]].
%37 citations counted in INSPIRE as of 19 Dec 2021

%\cite{Drepanou:2021jiv}
\bibitem{Drepanou:2021jiv}
N.~Drepanou, A.~Lymperis, E.~N.~Saridakis and K.~Yesmakhanova,
``Kaniadakis holographic dark energy,''
[arXiv:2109.09181 [gr-qc]].
%8 citations counted in INSPIRE as of 19 Dec 2021

\bibitem{Majhi:2017zao}
A.~Majhi,
``Non-extensive Statistical Mechanics and Black Hole Entropy From Quantum 
Geometry,''
Phys. Lett. B \textbf{775}, 32-36 (2017)
doi:10.1016/j.physletb.2017.10.043
[arXiv:1703.09355 [gr-qc]].

\bibitem{Mejrhit:2019oyi}
K.~Mejrhit and S.~E.~Ennadifi,
``Thermodynamics, stability and Hawking\textendash{}Page transition of 
black holes from non-extensive statistical mechanics in quantum geometry,''
Phys. Lett. B \textbf{794}, 45-49 (2019)
doi:10.1016/j.physletb.2019.03.055

\bibitem{Liu:2021dvj}
Y.~Liu,
``Non-extensive Statistical Mechanics and the Thermodynamic Stability of 	
FRW Universe,''
doi:10.1209/0295-5075/ac3f52
[arXiv:2112.15077 [gr-qc]].

%\cite{Nojiri:2022aof}
\bibitem{Nojiri:2022aof}
S.~Nojiri, S.~D.~Odintsov and V.~Faraoni,
``From nonextensive statistics and black hole entropy to the holographic dark universe,''
Phys. Rev. D \textbf{105} (2022) no.4, 044042
doi:10.1103/PhysRevD.105.044042
[arXiv:2201.02424 [gr-qc]].
%4 citations counted in INSPIRE as of 12 Mar 2022

\bibitem{Faraoni:2021nhi}
V.~Faraoni, A.~Giusti and B.~H.~Fahim,
``Spherical inhomogeneous solutions of Einstein and 
scalar\textendash{}tensor gravity: A map of the land,''
Phys. Rept. \textbf{925}, 1-58 (2021)
doi:10.1016/j.physrep.2021.04.003
[arXiv:2101.00266 [gr-qc]].

\bibitem{Heinzle:2001bk}
J.~M.~Heinzle and R.~Steinbauer,
``Remarks on the distributional Schwarzschild geometry,''
J. Math. Phys. \textbf{43}, 1493-1508 (2002)
doi:10.1063/1.1448684
[arXiv:gr-qc/0112047 [gr-qc]].

\bibitem{Steinbauer:2006qi}
R.~Steinbauer and J.~A.~Vickers,
``The Use of generalised functions and distributions in general 
relativity,''
Class. Quant. Grav. \textbf{23}, R91-R114 (2006)
doi:10.1088/0264-9381/23/10/R01
[arXiv:gr-qc/0603078 [gr-qc]].

\bibitem{Veraguth:2017uwp}
O.~J.~Veraguth and C.~H.~T.~Wang,
``Immirzi parameter without Immirzi ambiguity: Conformal loop 
quantization of scalar-tensor gravity,''
Phys. Rev. D \textbf{96}, no.8, 084011 (2017)
doi:10.1103/PhysRevD.96.084011
[arXiv:1705.09141 [gr-qc]].

\bibitem{Wang:2018bdg}
C.~H.~T.~Wang and D.~P.~F.~Rodrigues,
``Closing the gaps in quantum space and time: Conformally augmented gauge 
structure of gravitation,''
Phys. Rev. D \textbf{98}, no.12, 124041 (2018)
doi:10.1103/PhysRevD.98.124041
[arXiv:1810.01232 [gr-qc]].

\bibitem{Wang:2019ryx}
C.~H.~T.~Wang and M.~Stankiewicz,
``Quantization of time and the big bang via scale-invariant loop 
gravity,''
Phys. Lett. B \textbf{800}, 135106 (2020)
doi:10.1016/j.physletb.2019.135106
[arXiv:1910.03300 [gr-qc]].

%\cite{Nojiri:2017kex}
\bibitem{Nojiri:2017kex}
S.~Nojiri and S.~D.~Odintsov,
``Regular multihorizon black holes in modified gravity with nonlinear electrodynamics,''
Phys. Rev. D \textbf{96} (2017) no.10, 104008
doi:10.1103/PhysRevD.96.104008
[arXiv:1708.05226 [hep-th]].
%52 citations counted in INSPIRE as of 11 Mar 202

\end{thebibliography}
\end{document}